# Automatic Pill Reminder For Easy Supervision

*A. Jabeena.*
Associate Professor, School of Electronics Engineering
VIT University
Vellore, India
ajabeena@vit.ac.in

*Animesh Kumar Sahu*
School of Electronics Engineering
VIT University
Vellore, India
sahuanimesh94@gmail.com

*Rohit Roy*
School of Electronics Engineering
VIT University
Vellore, India
rohit.roy2014@gmail.com

*N.Sardar Basha*
Faculty
International Maritime College
Oman, Sultanate of Oman
sardar@imco.edu.com

*Abstract*— **In this paper we present a working model of an automatic pill reminder and dispenser setup that can alleviate irregularities in taking prescribed dosage of medicines at the right time dictated by the medical practitioner and switch from approaches predominantly dependent on human memory to automation with negligible supervision, thus relieving persons from error-prone tasks of giving wrong medicine at the wrong time in the wrong amount**.

## I. Introduction

Today, in-home 24-7 healthcare monitoring and supervisory facilities demand a great amount of money and human labor. This, added with the innate forgetfulness of the human mind can result in grave irregularities, often leading to negligence, critical situations and despair. Often we cannot comprehend the harm we inflict on our body by not taking pills at the right time, delaying intake or leaving it midway altogether, or even erroneously taking the wrong dosage. While automation and technology has helped in some of the key sectors to remove human error and achieve a desired level of efficiency, taking pills at the right time is still not seen as an area which could be synchronized by modern technology. Conventional pill bottles could be reconfigured into an automatic multi-pill reminder and dispenser for ease of operation and usability. With the help of simple microcontroller, we have attempted to create a model of a pillbox - an automated pill reminder and dispenser containing several compartments for keeping different types of pills such as tablets, capsules and suppositories having prescribed administration schedule and is a modern day alternate to a standard pill dispenser [6]. It uses a micro-controller to keep track of when a patient should take his/her medication [2]-[5]. It displays the time for the next medicine in a LCD screen and when the time arrives, it generates messages repeatedly, along with LED blinking indicating which compartment to open. When the patient opens a compartment, a sensor detects this and resets the light, alarm gets snoozed.

## II. Motivation

The motivation for building such a product came from the observing the elderly in the family having to suffer from the distress of missing out pills in absence of regular supervision[1].Public hospitals also display lack of supervision at times, which aggravates the situation. All these happenings motivated us to come up with an easy solution for the aged and the ailing.

## III. Related Work

Several different pillbox were available in the market. The cheapest one was the traditional pillbox, which contained seven boxes for seven different days of a week, costing around 200 rupees. However, user had to load the pills to the boxes every week. Mixing different pills in the same box would increase the risk of making mistakes. We also found another type of pillbox, which had the sound reminder, and was able to remind the user to take medicine at user specified time. However, the users still have to put different kinds of pills in the same box, and reload the boxes every week. Additionally, it could only remind the user to take pills once a day. The average costs of this type of pillbox were about 1000 INR, Therefore, we think it was necessary to build a cheap and functional smart Medicine Box that could bring more convenience for the user. [6] We then defined the specifications of our device based on the user needs. From the literature cited, the research proposed an idea of Smart Medicine Box [9]-[10] that will adapt the features of time tracking and alarm triggering Additionally, as compared to the existing system, It will remind the user to take medicine not for once per day but thrice per day along with that user does not need to refill the box every week.

## IV. MAJOR COMPONENTS USED

The project entails a simple electronic reminder system comprising of the following components:

Real time clock, Arduino Uno, Arduino IDE, IR sensors, Buzzer, GSM Module, LCD Display, Pill Box, connecting wires, jumper wires, and breadboard.

### A. Medicine Box

A rectangle box is divided into three equal sub-boxes where each sub-box contains a LED and buzzer is fitted onto the top of the box. The pill box model is as shown in Fig. 1.

### B. Arduino

Arduino Uno is a microcontroller board based on the ATmega328P (datasheet). It has 14 digital input/output pins (of which 6 can be used as PWM outputs), 6 analog inputs, a 16 MHz quartz crystal, a USB connection, a power jack, an ICSP header and a reset button. Arduino consists of both a physical programmable circuit board (often referred to as a microcontroller) and a piece of software, or IDE (Integrated Development Environment) that runs on your computer, used to write and upload computer code to the physical board.

### C. LCD

Liquid-crystal display (LCD) is a flat panel display, electronic visual display that uses the light modulating properties of liquid crystals. Liquid crystals do not emit light directly. In our project 16x4 LCD is used to display the information about pillbox such as the number of medicines in each sub-box to be consumed when the alarm rings.

### D. GSM

GSM (Global System for Mobile Communications, originally Groupe Spécial Mobile), is a standard developed by the European Telecommunications Standards Institute (ETSI).It was created to describe the protocols for second-generation (2G) digital cellular networks used by mobile phones and is now the default global standard for mobile communications – with over 90% market share, operating in over 219 countries and territories. The SIM800L module supports quad-band GSM/GPRS network, available for GPRS and SMS message data remote transmission.

### E. Real Time Clock Module

Real Time Clock (RTC) module uses the DS1307 to keep track of the current year, month, day as well as the current time. It includes small lithium coin cell battery that will run the RTC and can be accessed via the I2C protocol. In our project it used to set a specific time as per the patient required i.e. if the user wants to set 8.00 am as its morning medicine taking time then they can do with the help of this module.

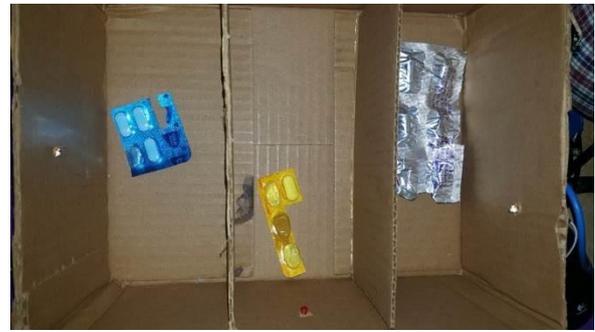

Fig. 1. Pill Box Subdivision

## V. PROPOSED MODEL

The setup consists of a small box divided into multiple compartments, each having a lid to open and an IR sensor attached to it. The box is connected to a real time clock, a microcontroller device Arduino Uno which processes the activities and accordingly displays the pill details and time of intake on the LCD attached to the box and a GSM module which sends message to the family physician or members in case the pill is not taken. The box consists of several compartments each having a pill for a definite time of the day. An electronic real time clock, with factory predetermined time interval, is automatically activated in sync with the pill intake timings. The real time clock will start beating and as it reaches the stipulated time of pill-intake, the buzzer will go on and message will be displayed regarding which pill to take and time to take each pill [7]-[8].

The pill dispenser may be preloaded by the patient himself or may be preloaded by someone assisting the patient once a day, thereby minimizing or totally eliminating the possible confusion as to when to take the prescribed medicine and what dosages to take.

*a)* Now if the person/user takes the pills, i.e. opens the lid, the IR Sensor attached to the lid will detect that the lid is opened and hence will send the output to Arduino which will stop the buzzer. This will be taken into the log registering the person has taken his medicine successfully.

*b)* In case the person fails to take the medicine or refuses to, the lid will not open and the buzzer will automatically stop after a preset time and will be put on snooze. If a person again misses the medicine, the output will be sent to the GSM module attached which in turn will send a message to the person reminding him that he has missed a pill. And if once again the person misses the pill, a message will be sent to family members.

The block diagram of the proposed model, process flow, front and side view of the model are as shown in Fig.2 ,Fig. 3 ,Fig. 4 and Fig. 5 respectively.

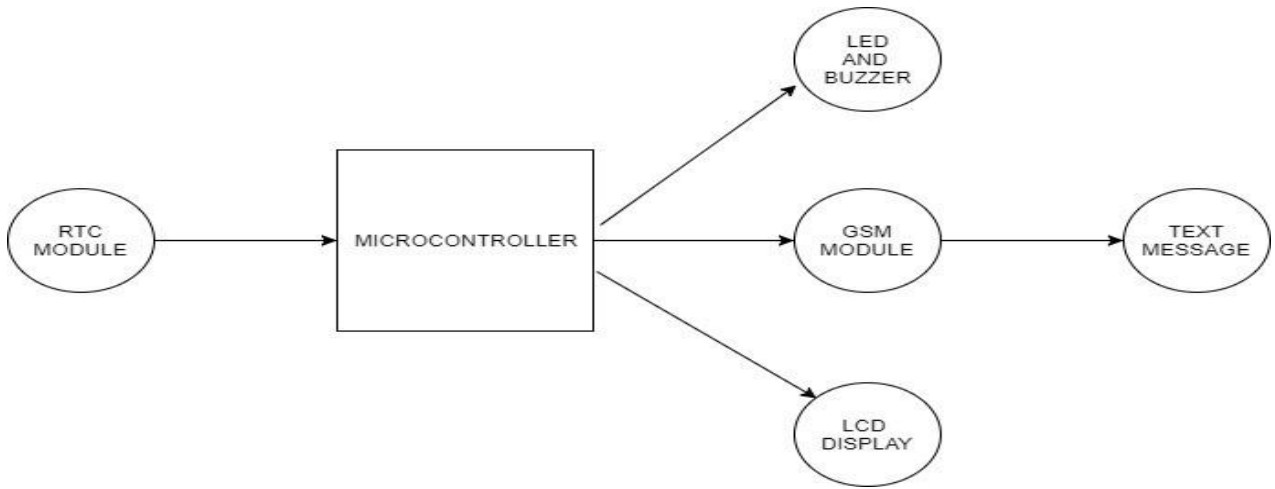

Fig. 2. Block Diagram of the Proposed Model

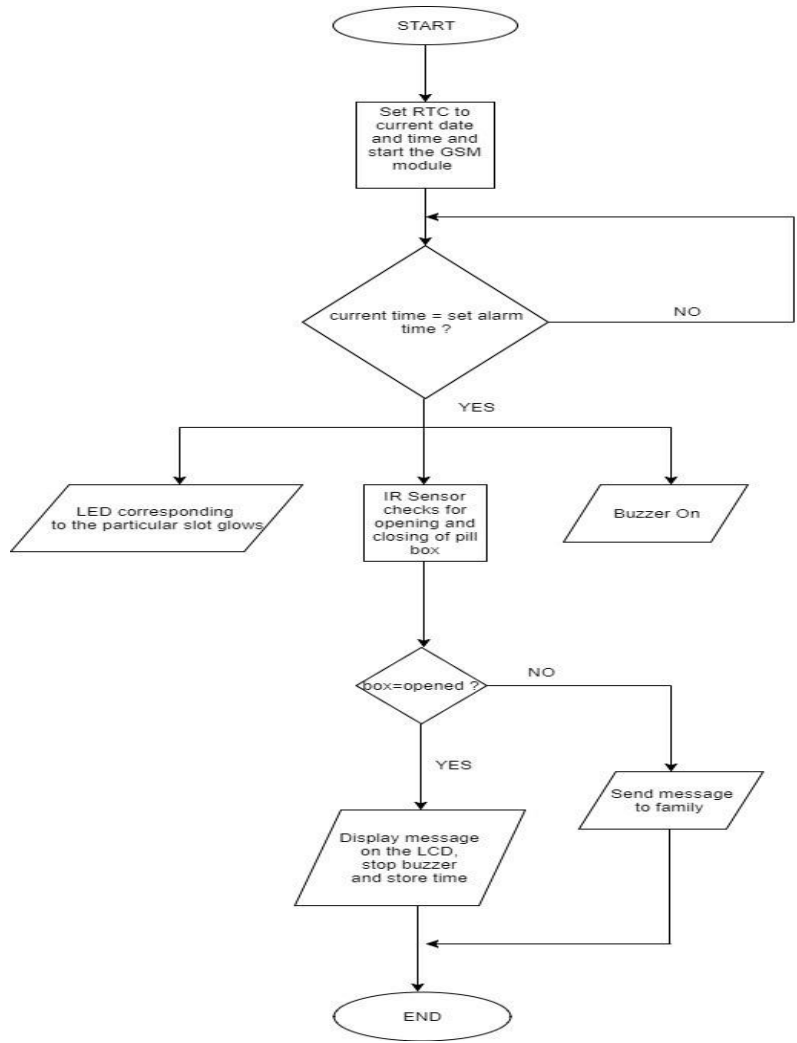

Fig. 3. Process Flow Schematic

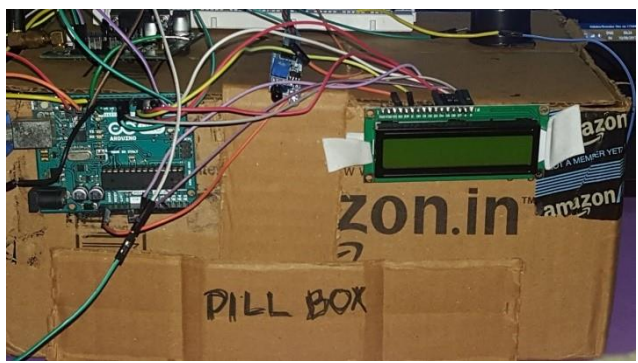

Fig. 4. Front End View of the Model

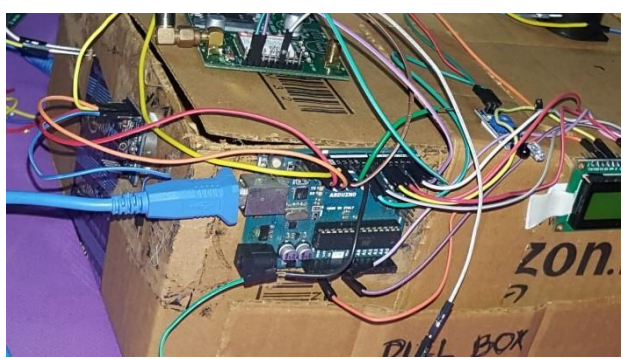

Fig. 5. Side View of the Model

## VI. RESULTS AND DISCUSSIONS

The device helps in keeping track of regular medical taking activities and reduces manual supervision and human effort. With simple circuitry and effort, the easy-to-use and cheap device comes as a boon for the young and the elderly, a simple solution for mothers for their adolescents, and caretakers for the aged and suffering. It can find its use in every household or hospital that has medical supervision problem and can be marketed as an efficient solution to us. The desired output results are shown in the Fig. 6 and Fig. 7.

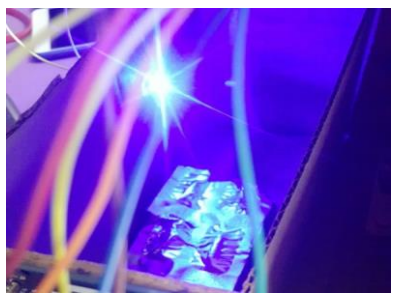

Fig. 6. Time to take the Medicine

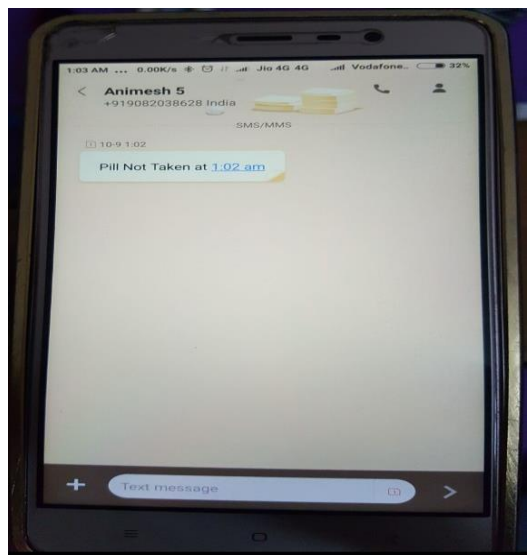

Fig. 7. Message If Pill Not Taken

## VII. CONCLUSION

In this paper a low-cost, useful model for an automatic pill reminder and disperser box has been designed using simple electronics applications. For easy detection and alert, a buzzer and a LCD display has been attached so that the person in concern takes his pill in time in the right quantity without personalized supervision. The device also records the time and date of taking pills as a useful database for future medical consultation. Family members are alerted, if pills are not taken on time. This easy-to-use device can be a convenient option for households where family members have work-hour compulsions or are compelled to keep a mistress for the member with medical complications.

REFERENCES

[1] MacLaughlin, Eric J., et al. "Assessing medication adherence in the elderly." Drugs & aging 22.3 (2005): 231-255.

[2] G. Eason, B. Noble, and I.N. Sneddon Lewis, Kermit E., and Arthur S. Roberts Jr. "Automatic pill dispenser and method of administering medical pills." U.S. Patent No. 4,573,606. 4 Mar. 1986.

[3] Shaw, Thomas J. "Automatic pill dispensing apparatus." U.S. Patent No. 5,609,268. 11 Mar. 1997.

[4] MacLaughlin, Eric J., et al. "Assessing medication adherence in the elderly." Drugs & aging 22.3 (2005): 231-255.

2005): 231-255.

[5] Fang, Kerry Y., Anthony J. Maeder, and Heidi Bjering. "Current trends in electronic medication reminders for self care." The Promise of New Technologies in an Age of New Health Challenges: Selected Papers from 5th Global Telehealth Conference 2016, Auckland, New Zealand, 1-2 November 2016. 2016.

[6] Billingsley, Luanne, and Ann Carruth. "Use of technology to promote effective medication adherence." The Journal of Continuing Education in Nursing 46.8 (2015): 340-342.

[7] Patel, Samir, et al. "Mobilizing your medications: an automated medication reminder application for mobile phones and hypertension medication adherence in a high-risk urban population." (2013): 630-639.

[8] Bai, Ying-Wen, and Ting-Hsuan Kuo. "Medication adherence by using a hybrid automatic reminder machine." Consumer Electronics (ICCE), 2016 IEEE International Conference on. IEEE, 2016.

[9] Balas, E. Andrew, et al. "Electronic communication with patients: evaluation of distance medicine technology." JAMA 278.2 (1997): 152-159.

[10] Heneghan, Carl J., P. Glasziou, and Rafael Perera. "Reminder packaging for improving adherence to self-administered long-term medications." Cochrane Database Syst Rev 1 (2006).

AUTHOR PROFILE

**A. Jabeena** is an Associate Professor in School of Electronics Engineering, VIT University, Vellore, Tamilnadu , India. Her research interest includes Optical communication, free space and Visible light communication, Embedded Technology.

**Animesh Kumar Sahu** and **Rohit Roy** are B. Tech, ECE students from School of Electronics Engineering, VIT University, Vellore, Tamil Nadu, India.

**N. Sardar Basha** is a faculty in the International Maritime College Oman, Sultanate of Oman.